\documentclass{ws-p8-50x6-00}

\def\ifmath#1{\relax\ifmmode #1\else $#1$\fi}%

\def\re{\ifmath{{\mathrm{e}}}}
\def\rK{\ifmath{{\mathrm{K}}}}
\def\rL{\ifmath{{\mathrm{L}}}}
\def\rp{\ifmath{{\mathrm{p}}}}

\def\rS{\ifmath{{\mathrm{S}}}}

\def\cm{\ifmath{{\mathrm{cm}}}}

\def\lab{\ifmath{{\mathrm{lab}}}}
\def\max{\ifmath{{\mathrm{max}}}}

\begin{document}

\title{Extensive Air Shower Simulations with {\it CORSIKA} and
the Influence of High-Energy Hadronic Interaction Models}

\author{D. Heck $~~$ for the {\bf KASCADE} Collaboration}
\address{Institut f{\"u}r Kernphysik, Forschungszentrum Karlsruhe,
D-76021 Karlsruhe, Germany\\E-mail:heck@ik3.fzk.de}

\maketitle

\abstracts{
When high-energy cosmic rays ($\gamma$'s, protons, or heavy nuclei) 
impinge onto the Earth's atmosphere, they interact at high 
altitude with the air nuclei as targets.
By repeated interaction of the secondaries an `extensive
air shower' (EAS) is generated with huge particle numbers
in the maximum of the shower development.
Such cascades are quantitatively simulated by the Monte Carlo computer
program CORSIKA. The most important uncertainties in simulations
arise from modeling of high-energy hadronic interactions:
a) The inelastic hadron-air cross sections.
b) The energies occurring in EAS may extend far above the energies
available in man-made accelerators, and when extrapolating towards
higher energies one has to rely on theoretical guidelines.
c) In collider experiments which are used to adjust the
interaction models the very forward particles are not accessible,
but just those particles carry most of the
hadronic energy, and in the EAS development they transport a
large energy fraction down into the atmosphere.\\
CORSIKA is coupled alternatively with 6
high-energy hadronic interaction codes (DPMJET, HDPM, {\sc neXus}, 
QGSJET, SIBYLL, VENUS).
The influence of those interaction models on observables
of simulated EAS is discussed.
}

\section{Introduction}

CORSIKA ({\bf CO}smic {\bf R}ay {\bf SI}mulation for {\bf KA}scade) is
a detailed Monte Carlo program to study the evolution of extensive air
showers (EAS) in the atmosphere initiated by various cosmic ray particles.
Originally, it was designed to perform simulations for the KASCADE
experiment\cite{kascade} at Karlsruhe and has been
refined considerably since its first version in 1989.
Meanwhile, it has developed into a tool that is used 
for more than 30 experiments
worldwide. The prediction of particle energy spectra, densities or
arrival times to be observed in EAS experiments is a well suited
application of CORSIKA.
A detailed description of the physics incorporated in CORSIKA
is given in Ref.\cite{corsika_phys}, technical details on the handling
of the program are described elsewhere.\cite{corsika_tech}

\section{EAS Environment Parameters and Particle Transport}

To simulate the evolution of EAS global parameters have to be specified:
The Earth's magnetic field affecting the movement of charged particles
as well as the atmospheric model to be employed in the simulation
depend on the geographic location. 
CORSIKA provides several atmospheric parameter sets covering the complete
climatical and seasonal influence from tropical, subtropical, and
mid-latitude regions to the South pole (4 seasons).
The composition of air is adopted to 78.1\% N$_{2}$, 21.0\%
O$_{2}$ and 0.9\% Ar (volume fractions).

Within CORSIKA, various particles are followed: Besides
$\gamma$-rays the leptons $\re^{\pm}$ and $\mu^{\pm}$ ($\nu_{\re}$ 
and $\nu_{\mu}$ 
optionally), the mesons $\pi^{o}$,
$\pi^{\pm}$, $\rK^{o}_{\rS/\rL}$, $\rK^{\pm}$ and $\eta$, the nucleons,
the strange baryons with strangeness $\left|S\right| \le 3$, and the
corresponding anti-baryons as well as nuclei with $A \le 60$ are treated. 
Mesonic ($\rho, \omega, \rK^{*}$) and baryonic 
($\Delta$) resonances are decaying without transport. 

For each particle its transportation range is estimated. 
For instable particles both ranges for interaction (determined by 
cross section) and decay (limited by lifetime) are evaluated 
independently and the shorter one determines the fate of the 
particle at the end of its range.
In decays, all branches down to the 1 \% level are considered with 
correct kinematics in the 3-body decays.
In the range determination of decaying charged particles, the 
ionization energy loss is considered, which especially affects 
muons at energies below $\approx 10$~GeV because of
their long lifetime and low interaction cross section.
This treatment slightly favors the decay of charged pions 
- the main source of muons - at the expense of pion-induced 
interactions, dependent on energy and height of pion origin. 
During transport, the deflection of charged particles by 
the Earth's magnetic field is considered.

In CORSIKA, electrons and gammas are treated in a tailored 
version of the EGS4 code\cite{egs} and/or, less detailed but 
much faster, by analytical NKG-formulas.\cite{nkg}
The electromagnetic interactions are believed to be treated 
correctly even at the highest energies by taking into 
account the Landau-Pomeranchuk-Migdal effect. 
Cherenkov photons may be generated optionally.
Hadronic interactions with energies $E_{\lab} \le 80$~GeV are 
modeled by the GHEISHA code\cite{gheisha} or, alternatively, by 
the UrQMD model.\cite{urqmd}

\section{Hadronic Interaction Models}

\begin{table}[t]
\caption{Essential features of hadronic interaction models.}
\scriptsize{
\begin{center}
\begin{tabular}{|l|cccccc|}
\hline
                &       &        &      &      &      &   \\[-1mm]
Model           & VENUS &neXus   &QGSJET&DPMJET&SIBYLL&HDPM\\
Version         & 4.12  & 2      &      & II.4/II.5 & 1.6/2.1  &    \\
\hline
                &       &        &      &      &      &   \\[-2mm]
Gribov-Regge    &   +   &   +    &  +   &   +  &      &    \\
Mini-Jets       &       &   +    &  +   &   +  &   +  &    \\
Sec. Interactions & +   &   +    &      &      &      &    \\
N-N interaction &   +   &   +    &  +   &   +  &      &    \\
Superposition   &       &        &      &      &   +  &  + \\
Max. Energy (GeV)&2$\cdot$10$^7$&2$\cdot$10$^8$&$>$10$^{11}$&$>$10$^{11}$&$>$10$^{11}$&10$^8$\\
Memory (Mbyte)& 21    &    101 &  10  &  52  &   9  &  8 \\
CPU-Time$^1$ (min)&4.5&$\approx$100&1.0  & 3.5  & 0.75 & 1.0\\
\hline
\multicolumn{7}{l}{$~$}\\[-2mm]
\multicolumn{7}{l}{$^1$ for showers with primary p, $E_0=10^{15}$~eV, 
vertical, $E_h$, $E_{\mu} \ge 0.3$~GeV, 110~m a.s.l.,}\\
\multicolumn{7}{l}{$^{~}$ NKG option, DEC 3000/600 AXP ($175$~MHz)}\\
  \end{tabular}  
\end{center}
}
\label{tab-models}
\end{table}
 
The largest uncertainties in numerical simulation of EAS with primary 
energies above some TeV are induced by the models which describe
the hadronic interactions.
Especially when extrapolating to the highest energies, where the collision 
energies exceed those accessible by accelerators, one has to rely on 
theoretical guidelines to describe the interactions.
Additional uncertainties stem from the fact that just those 
interaction products emitted at small angles into the extreme forward 
direction carry away the largest energy fraction, 
but in collider experiments those particles disappear in the beam pipe 
without being observable.
In the development of EAS, such particles are responsible for
transporting the energy down into the atmosphere.

To study the influence of models on the uncertainties of EAS observables 
and on their correlations,
6 different hadronic interaction codes DPMJET\cite{dpmjet}, 
HDPM\cite{corsika_phys}, {\sc neXus}\cite{nexus}, QGSJET\cite{qgsjet}, 
SIBYLL\cite{sibyll,sibyll_new}, and VENUS\cite{venus} have been  
coupled with CORSIKA. 
They describe the hadronic interactions at
energies $E_{\lab} \ge 80$~GeV.
Their basic features are summarized in Table \ref{tab-models}.

\section{Cross Sections}

The longitudinal development of hadronic EAS depends 
crucially on the inelastic hadron-air cross section. 
Lower cross sections elongate, higher ones 
shorten the longitudinal development.
\begin{figure}[h]
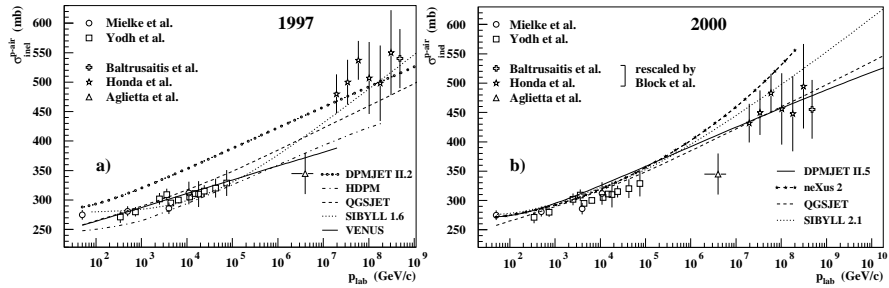

\hspace{-1.5mm}{
 \label{fig-sigma}
  \epsfig{file=sigmod_p_air_bw_old.epsf,width=57mm}
 \hspace{-3.5mm}{
  \epsfig{file=sigmod_p_air_sib21_nex2_dpj25_bw.epsf,width=64mm}
 }
}
\vspace{-6mm}{
 \caption{Inelastic p-air cross sections in the years 1997  and 2000 
     (Refs.\protect\cite{mielke,yodh,baltrusaitis,honda,aglietta,block}).}
}
\end{figure}
Fortunately, the situation has improved in the last 3 years,
as demonstrated in Fig.$~$1.     
The cross section differences between the models have shrunk 
from 80~mb to today 20~mb in the region at a few PeV.
Extrapolations to the highest energies above $10^{10}$~GeV 
become much more realistic to make predictions for the Auger 
experiment.\cite{dova}

\section{Comparison with Collider Data}

All models are adjusted to experimental data wherever available. 
Especially $\rp\overline{\rp}$ collider data should be reproduced with 
respect to e.g. the increase of average charged particle multiplicity with 
increasing energy, the spread of the number of emitted charged particles 
around the mean value according to a negative binomial distribution 
as shown in  Fig.~2a, 
and the pseudorapidity distribution of emitted charged particles 
shown together with recent experimental values\cite{harr} at 
$E_{\cm} = 630$~GeV in Fig.~2b.
The more modern models\cite{dpmjet,nexus,qgsjet,sibyll_new} displayed here
show good agreement with experimental data.
A complete comparison of the older models can be found in Ref.\cite{comparison}.
\begin{figure}[b]
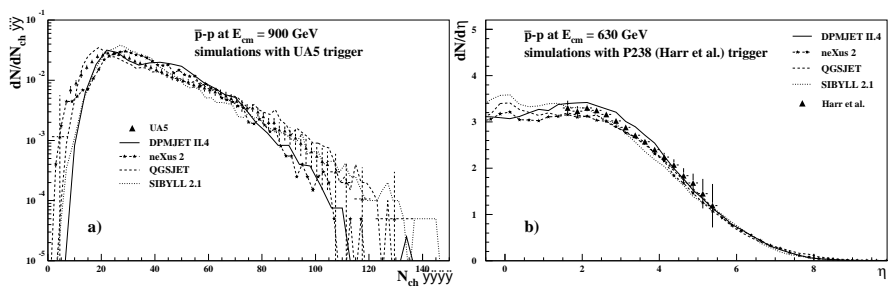

\hspace{-2mm}
\parbox{57mm} {
 \label{fig-collider}
 \epsfig{file=ppn900_ua5_bw.epsf,width=61mm}
}
\parbox{56mm} {
 \epsfig{file=prap630_harr_bw.epsf,width=61mm}
}
\vspace{-2mm}{
 \caption{
   Model adjustments to experimental values\protect\cite{ua5,harr}
             of $\rp\overline{\rp}$ collisions:
   Charged particle multiplicities at $\sqrt s = 900$~GeV (a),
   Charged particle pseudorapidity distributions at 
             $\sqrt s = 630$~GeV (b).  
 }
}
\end{figure}

For some parameters, 
the CORSIKA interaction tests revealed differences responsible for 
deviating properties of EAS simulated with different models.
The most frequent hadronic interaction within the development of an 
EAS cascade is the collision of a charged pion with a $^{14}$N nucleus. 
Therefore, we compare in Fig.~3a
the model predictions of the average charged particle 
multiplicity produced by this reaction as a function of energy. 
In general, the higher the multiplicity, the more energy is 
dissipated into the 
production of secondaries, which gives rise to a shorter EAS 
longitudinal distribution. 
The highest multiplicities are observed for QGSJET.
\begin{figure}[t]
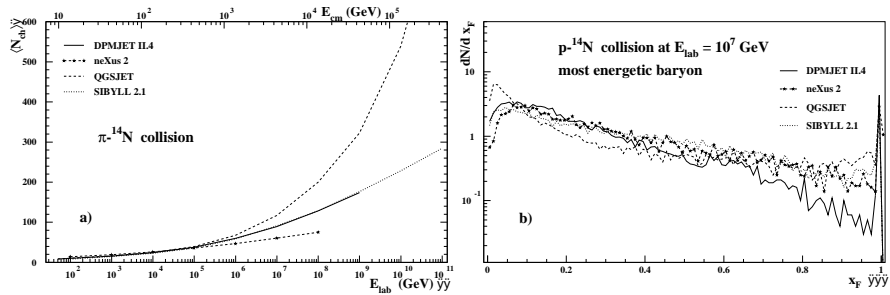

\hspace{-2mm}
\parbox{57mm} {
 \label{fig-predict}
 \epsfig{file=piNn3_bw.epsf,width=61mm}
}
\parbox{56mm} {
 \epsfig{file=pNxf1e16_detail_bw.epsf,width=61mm}
}
\vspace{-2mm}{
 \caption{
  Model predictions for hadron-$^{14}$N collisions:
  Average charged particle multiplicity as a function of
             energy for $\pi$-$^{14}$N collisions (a),
  energy fraction (Feynman-x) distribution of the most energetic baryon from
             p-$^{14}$N collisions at $E_{\lab} = 10^7$~GeV (b).
 }
}
\end{figure}
In those events with many secondary particles only a small energy
fraction is left for the relic of the projectile. 
This is demonstrated in Fig.~3b,
which shows the Feynman-x distribution of the most energetic 
baryon from p-$^{14}$N collisions.
In the range $0.1 < {\rm x_{_F}} <0.4$, corresponding to a 
moderate energy transfer to secondary particles, QGSJET exhibits
the lowest rate.

Of special interest is the behavior of the 
Feynman-x 
distribution of Fig.$~$3b for values ${\rm x_{_F}} \ge 0.85$. 
This region is governed by diffractive interactions. 
In diffractive collisions the projectile looses only a small energy 
fraction, the largest energy portion is transported deeper into 
the atmosphere with the projectile remainder.
It must be pointed out that describing correctly the diffractive 
phenomena is of great importance for the hadron rates which 
are observed at KASCADE level.\cite{risse}
The large differences between the models reflect the lack of
experimental data in the forward region.

\section{Results of EAS Simulations}

Only new or recently revised models\cite{dpmjet,nexus,qgsjet,sibyll_new} 
are regarded here. 
A more detailed comparison of the 
models is given in Ref.\cite{comparison,gransasso}. 
All shower properties which can be measured with detectors placed 
at  observation level more or less strongly depend on the 
`age' of EAS development. 
As mentioned above, the inelastic hadronic cross sections 
as well as the `inelasticity' (the fraction of energy carried
away by secondary particles) of  hadronic interactions are the 
essential quantities which determine the elongation (slow aging) 
or shortening (fast aging) of the EAS development. 

In  Fig.~4a the number of electrons (incl. positrons) is shown as a 
function of atmospheric depth for vertical showers induced by $protons$ 
or $Fe$-nuclei with energies of $E_0 = 10^{14}$ and $10^{15}$~eV. 
\begin{figure}[t]
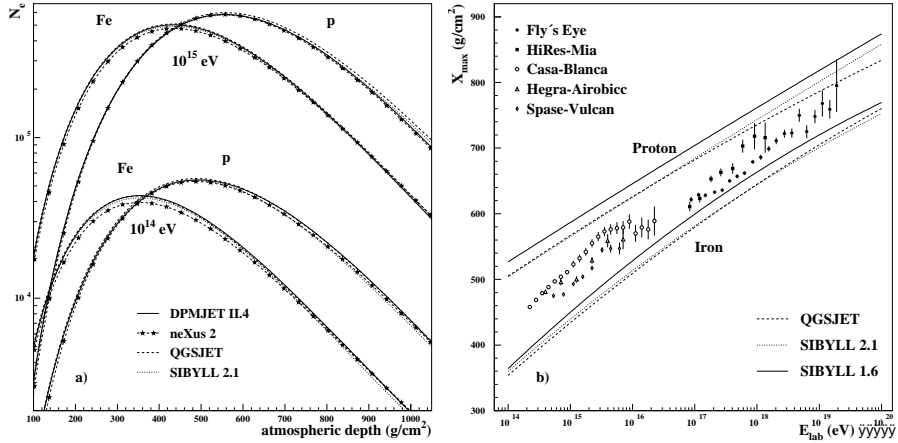

\hspace{-3.5mm}{
  \label{fig-results}
  \epsfig{file=longi2_nex2_sib21_bw.epsf,width=60.5mm}
  \epsfig{file=xmax_smooth.epsf,width=60.5mm}
}
\vspace{-3mm}
\caption{Longitudinal shower development for proton and iron induced
         vertical showers at primary energies of $E_0 = 10^{14}$~eV 
         and $10^{15}$~eV (a).
         Depth of shower maximum X$_{\max}$ as a function of energy as
         predicted by QGSJET and two versions of SIBYLL, together with 
         measurement points of 
         experiments\protect\cite{flys_eye,hires,casa_blanca,hegra,spase} (b).}
\vspace{-0.5mm}
\end{figure}
The number (averaged over 500 simulated showers each)
of electrons at sea level (1036~g/cm$^{2}$) differs between
the models by max. 14 \% ($proton$) resp. 3 \% ($Fe$) at $10^{15}$~eV,
becoming even smaller at lower energies.
The differences have shrunk by more than a factor of 3 
with respect to a 1997 comparison\cite{gransasso} of the older models.

A quantity measured by several experiments is the depth $X_{\max}$ 
(expressed in g/cm$^{2}$) of the maximum number of charged particles 
within the EAS development. 
The $X_{\max}$-value is sensitive to the primary particle type and may be 
used to determine the (energy dependent) cosmic ray mass composition.
Fig.~4b gives a survey of several experiments, which use different 
techniques (fluorescence\cite{flys_eye,hires} = filled symbols,  
Cherenkov\cite{casa_blanca,hegra,spase} = open symbols) 
to determine $X_{\max}$ at various energies, extending over 5 decades.

The model predictions of Fig.$~$4b are derived from averages 
over 500 vertical $proton$-induced showers (resp. 200 $Fe$-induced showers) 
for each of the 13 equidistant reference energies. 
The uncertainties of the mean $X_{\max}$ are dominated by shower fluctuations 
and range from 5~g/cm$^2$ at $10^{14}$~eV to 3~g/cm$^2$ at 
$10^{20}$~eV for $proton$ EAS and amount to about the half for $Fe$.
The mean $X_{\max}$-values are approximated by quadratic expressions of 
the form
\begin{displaymath}
X_{\max} = a + b \cdot \lg E + c \cdot (\lg E)^2\ , 
\end{displaymath}
which follow the simulated mean values within the error bars.

It must be emphasized that none of the models predicts a linear 
relation as is often assumed in oversimplified arguments.
The elongation rate per energy decade decreases from 
c.~70~g/cm$^2$ at $10^{14}$~eV  to c.~50~g/cm$^2$ at 
$10^{20}$~eV for $proton$-induced showers and shows a similar 
reduction from c.~80 to c.~60~g/cm$^2$ for $Fe$-induced showers.
Remarkable is the good agreement between QGSJET and 
SIBYLL 2.1 up to $10^{18}$~eV. 
At higher energies, SIBYLL predicts 
a larger  $X_{\max}$-separation between $proton$ and $Fe$ showers.
The new SIBYLL version 2.1 generally  reveals 
$X_{\max}$-values reduced by $\approx 20$~g/cm$^2$ relative to the older
version 1.6.

\section{Conclusion}

In the last 3 years large progress is attained in EAS simulations.
By the reevaluation of inelastic proton-air cross sections,
a considerable agreement is achieved now up to the $10^7$~GeV range.
A clear trend of convergence between different hadronic interaction
models is obvious for primary energies up to the $10^6$~GeV range.  
Despite its age, presently QGSJET still shows the best, though not in all 
respects satisfying agreement with a variety of experimental 
results.\cite{risse,kampert}
The new SIBYLL version\cite{sibyll_new} and the 
more modern ideas realized in {\sc neXus}\cite{nexus}
should enable a more consistent and reliable 
extrapolation especially to the highest energies.
The announced\cite{dpmjet3} version III of DPMJET  
still has to demonstrate its improved quality.

\section*{Acknowledgments}

Many thanks go to the authors of the hadronic
interaction models for their help to get their programs
running and for their advice in coupling the programs with CORSIKA.
I am indebted to  R. Engel for making available the new SIBYLL 
version 2.1.
The partial support of this work by the British German Academic Research  
Collaboration (Grant 313/ARC-lk) is acknowledged.


\end{document}
